\documentclass[12pt,preprint]{aastex}
\usepackage{natbib}
\usepackage{amssymb,amsmath}

\shorttitle{ GRB 150910A }
\shortauthors{Xie et al. }

\begin{document}

\title{Early Optical Observations of GRB 150910A: Bright Jet Optical Afterglow and X-ray Dipole Radiation from a Magnetar Central Engine}

\author{Lang Xie\altaffilmark{1,2}, Xiang-Gao Wang\altaffilmark{1,2}, Song-Mei Qin\altaffilmark{3}, WeiKang Zheng\altaffilmark{4}, Alexei V. Filippenko\altaffilmark{4,5}, Long Li\altaffilmark{1,2}, Tian-Ci Zheng\altaffilmark{1,2}, Le Zou\altaffilmark{1,2}, Da-Bin Lin\altaffilmark{1,2}, Yinan Zhu\altaffilmark{6}, Heechan Yuk\altaffilmark{4,7}, Rui-Jing Lu\altaffilmark{1,2}, and En-Wei Liang\altaffilmark{1,2}}

\altaffiltext{1}{Guangxi Key Laboratory for Relativistic Astrophysics, School of Physical Science and Technology, Guangxi University, Nanning 530004, China; wangxg@gxu.edu.cn, lew@gxu.edu.cn}
\altaffiltext{2}{GXU-NAOC Center for Astrophysics and Space Sciences, Nanning 530004, China}
\altaffiltext{3}{Mathematics and Physics Section, Guangxi University of Chinese Medicine, Nanning 53001,China}
\altaffiltext{4} {Department of Astronomy, University of California, Berkeley, CA, 94720-3411, USA; weikang@berkeley.edu, afilippenko@berkeley.edu}
\altaffiltext{5} {Miller Senior Fellow, Miller Institute for Basic Research in Science, University of California, Berkeley, CA 94720, USA}
\altaffiltext{6}{School of Physics and Astronomy, Sun Yat-sen University, Zhuhai 519082, China}
\altaffiltext{7}{Homer L. Dodge Department of Physics and Astronomy, University of Oklahoma, 440 W. Brooks St. Norman, OK 73019, USA}

\begin{abstract}
  Gamma-ray burst (GRB) 150910A was detected by {\it Swift}/BAT, and then rapidly observed by {\it Swift}/XRT, {\it Swift}/UVOT, and ground-based telescopes. We report Lick Observatory spectroscopic and photometric observations of GRB~150910A, and we investigate the physical origins of both the optical and X-ray afterglows, incorporating data obtained with BAT and XRT. The light curves show that the jet emission episode lasts $\sim 360$~s with a sharp pulse from BAT to XRT (Episode I). In Episode II, the optical emission has a smooth onset bump followed by a normal decay ($\alpha_{\rm R,2} \approx -1.36$), as predicted in the standard external shock model, while the X-ray emission exhibits a plateau ($\alpha_{\rm X,1} \approx -0.36$) followed by a steep decay ($\alpha_{\rm X,2} \approx -2.12$). The light curves show obvious chromatic behavior with an excess in the X-ray flux. Our results suggest that GRB 150910A is an unusual GRB driven by a newly-born magnetar with its extremely energetic magnetic dipole (MD) wind in Episode II, which overwhelmingly dominates the observed early X-ray plateau. The radiative efficiency of the jet prompt emission is $\eta_{\gamma} \approx 11\%$. The MD wind emission was detected in both the BAT and XRT bands, making it the brightest among the current sample of MD winds seen by XRT. We infer the initial spin period ($P_0$) and the surface polar cap magnetic field strength ($B_p$) of the magnetar as $1.02 \times 10^{15}~{\rm G} \leq B_{p} \leq 1.80 \times 10^{15}~{\rm G}$ and 1~ms $\leq P_{0}v\leq 1.77$~ms, and the radiative efficiency of the wind is $\eta_w \geq 32\%$.
\end{abstract}


\section{ Introduction }
Gamma-ray bursts (GRBs) are the most luminous explosions in the Universe. Long-duration GRBs have been proposed to originate from core collapse of massive stars (e.g., Woosley 1993; Paczy\`{n}ski 1998; MacFadyen \& Woosley 1999; Kumar \& Zhang 2015; Dai et al. 2017; M\'{e}sz\'{a}ros et al. 2019). The collapse produces a rapidly spinning and strongly magnetized neutron star (millisecond magnetar) or a black hole.

In the millisecond magnetar scenario as the central engine of long GRBs, the magnetar could lose its rotational energy to produce a Poynting-flux-dominated outflow and power the GRB ejecta, with $E_{\rm rot} = (1/2) I \Omega_{0}^{2} \approx (2 \times 10^{52})~M_{1.4} R_6^2 P_{0,-3}^{-2}$ erg, where $I$ is the magnetar's moment of inertia, $\Omega_0 = 2\pi/P_0$ is its initial angular frequency, and $M_{1.4} = M/(1.4~M_\odot)$. During the period of jet magnetic dissipation, or shock collision, it could produce the prompt gamma-ray emission of GRBs. The residual rotational energy may generate a steady wind to produce a plateau-like phase in the early afterglow (i.e., X-ray plateau). After the characteristic spin-down timescale $\tau$ of the magnetar, its radiation luminosity evolves as $L\propto(1+t/\tau)^{\alpha}$ (e.g., Dai \& Lu 1998; Zhang \& M\'{e}sz\'{a}ros 2001; Liang et al. 2007; Troja et al. 2007; Metzger \& Piro 2014; L\"{u} \& Zhang 2014; Lasky \& Glampedakis 2016; Du et al. 2016; Chen et al. 2017), where $\alpha$ is respectively $-1$ and $-2$ in the gravitational wave (GW) and magnetic dipole (MD) radiation dominated scenarios (L\"{u} et al. 2018), and $\alpha < -3$ indicates that the magnetar may have collapsed to a black hole prior to spin-down.

Observed properties of GRBs and their early-time afterglows indicate the different structures for the central engine. Analyses of GRB light curves based on large samples (Nousek et al. 2006; O'Brien et al. 2006; Liang et al. 2007; L\"{u} \& Zhang 2014; L\"{u} et al. 2015, Wang et al. 2015; Wang et al. 2018) show that a significant fraction of X-ray afterglow light curves share common plateau features, and some exhibit rapid decay with $\alpha \leq -2$ (e.g., Troja et al. 2007; Liang et al. 2007; Lyons et al. 2010; L\"{u} \& Zhang 2014). Zou et al. (2019) found that the jet and MD wind radiation can be separated in a fraction of \emph{Swift} GRBs, also indicating that the shallow-decaying segment observed in the early-time X-ray afterglow light curves may be dominated by the MD radiation wind of a newly-born magnetar, which may serve as central engine of these GRBs.

GRB~150910A is an interesting GRB with an X-ray plateau in the early-time afterglow light curves,
which apparently exceeds the predictions of standard external shock models.
We suggest that the phase of prompt gamma-ray emission may be from jet radiation, and the X-ray plateau phase is mainly due to energy injection from the MD wind radiation of a millisecond magnetar in its early spin-down stage. The smooth onset feature observed in the optical afterglow light curves may be dominated by jet afterglow.

This paper reports our observations of a very bright optical afterglow of GRB 150910A and detailed modeling of the optical and X-ray afterglow light curves. Our observations and data analysis are presented in \S 2 and \S 3, respectively. Analysis of the jet properties and constraints on the central engine are presented in \S 4. A discussion of the results is given in \S 5, and \S 6 summarizes our conclusions. We assume a concordance cosmology of $H_0=69.6 ~\rm km \ s^{-1}~Mpc^{-1}$, $\rm \Omega_M = 0.286$, and $\Omega_\Lambda =0.714$ throughout the paper.

\section{Observations and Data Reduction}
The Burst Alert Telescope (BAT) onboard the \emph{Neil Gehrels Swift Observatory} (\emph{Swift}; Gehrels et al. 2004) triggered GRB~150910A at 09:04:48 (UT dates are used throughout this paper) on September 10, 2015 ($T_0$) in image mode (Pagani et al. 2015). Typical of image-triggered GRBs (such as GRB 060218; Campana et al. 2006), its real-time light curve shows as long-lasting flickering. The XRT and the Ultraviolet-Optical Telescope (UVOT) onboard {\it Swift} began observing the X-ray and optical afterglows 145~s and 153~s (respectively) after the BAT trigger (Pagani et al. 2015).

The bright optical counterpart of GRB~150910A was detected  by several ground-based telescopes, such as the 1-m telescope located at Nanshan, Xinjiang (Xu et al. 2015), the 10.4-m GTC  (Pagani et al. 2015), the Russian-SAO RAS 1-m telescope (Moskvitin et al. 2015), the 2.2-m MPG telescope at ESO La Silla Observatory (Schmidl et al. 2015), the Nordic Optical Telescope (Cano et al. 2015), the 1-m telescope of Tien Shan Astronomical Observatory (Mazaeva et al. 2015), and the Palomar 60-inch (P60) robotic telescope (Perley et al. 2015).

Our optical follow-up campaign of GRB~150910A was carried out using the
0.76-m Katzman Automatic Imaging Telescope (KAIT) at Lick Observatory, beginning at
$\sim T_0+1000$~s and ending $\sim 1.75$~hr after the {\em Swift}/BAT
trigger time (Zheng \& Filippenko 2015). The optical counterpart was
clearly detected in the $V$, $R$ and $Clear$ (close to $R$; see Li et al. 2003) bands.
KAIT data were reduced using our image-reduction pipeline (Ganeshalingam et al. 2010;
Stahl et al. 2019). Point-spread-function photometry was applied using DAOPHOT
(Stetson 1987) from the
IDL Astronomy User's Library\footnote{http://idlastro.gsfc.nasa.gov/}.
The multiband data were calibrated to local
Pan-STARRS1\footnote{http://archive.stsci.edu/panstarrs/search.php}
stars, whose magnitudes were transformed into the Landolt (1992) system using the empirical
prescription presented by Tonry et al. (2012, Eq. 6).

Additional photometric data were obtained with the 1-m Nickel telescope at
Lick Observatory during the second night, $\sim 0.911$ days after the trigger,
with an exposure time of $5 \times 600$~s in the $R$ band. The optical counterpart
was detected in the coadded image and was measured with the method above.
The afterglow light curves of GRB 150910A are shown in Figure \ref{fig-LCs}.

\begin{figure}
\centering
\includegraphics[angle=0,scale=0.5]{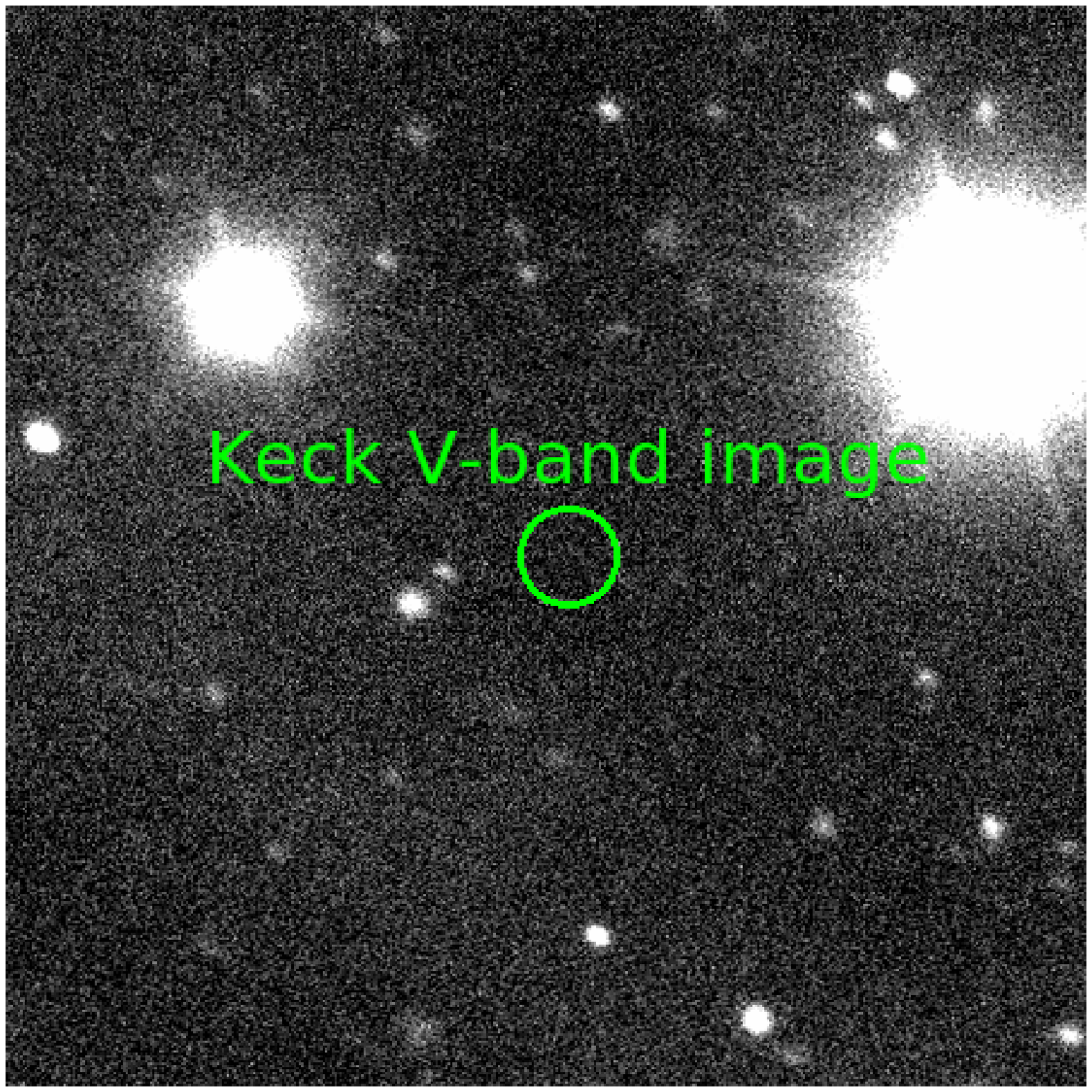}
\includegraphics[angle=0,scale=0.5]{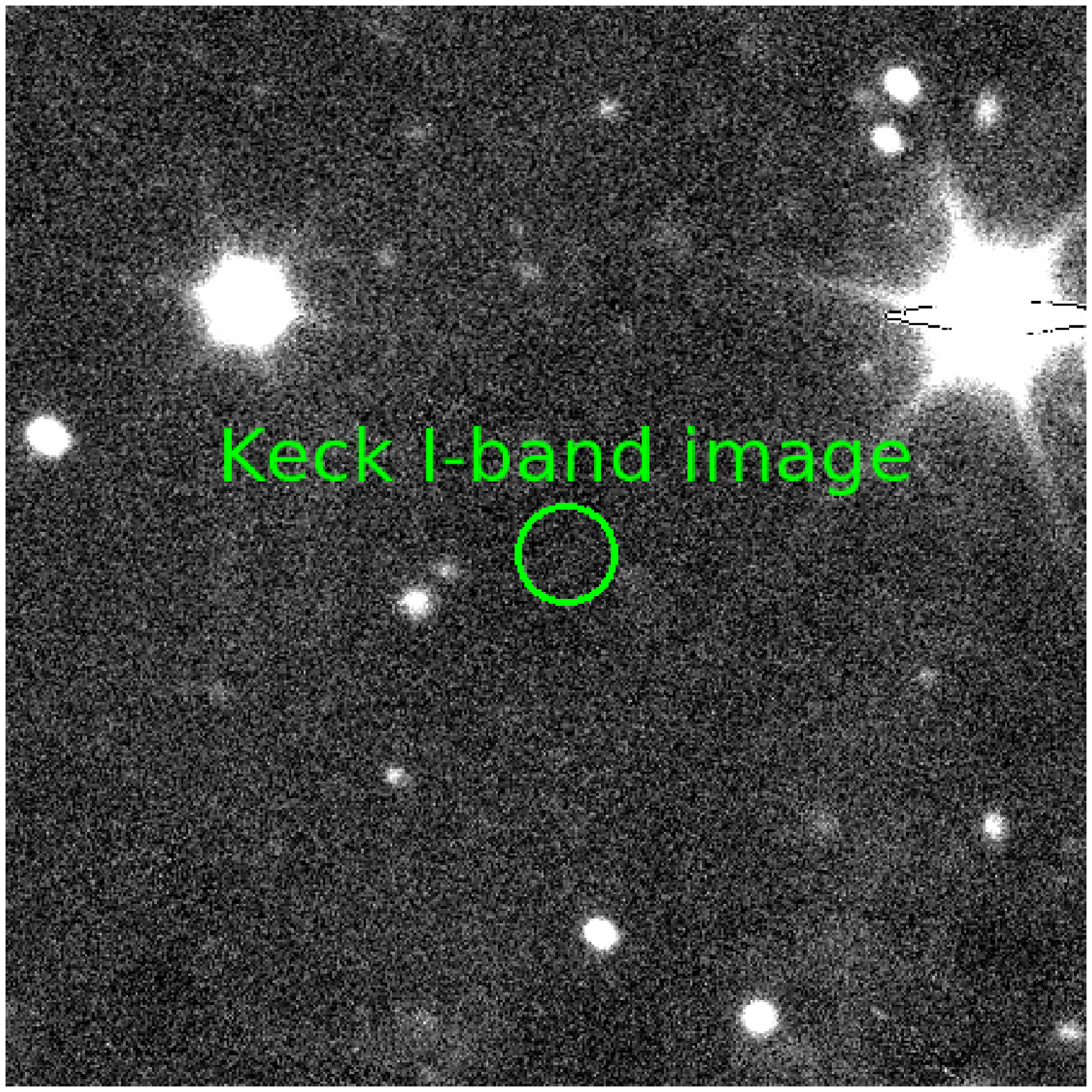}
\caption{Keck LRIS images of GRB150910A in the $V$ (left) and $I$ (right) bands taken
         on 2015~October~10. The optical counterpart was not detected; its
         position is marked with a green circle.}
\label{Fig_keck_VI_image}
\end{figure}

We also obtained a late-time deep image of the site of GRB~150910A
with the Low-Resolution Imaging Spectrometer
(LRIS; Oke et al. 1995) mounted on the 10-m Keck~I telescope on October~10, 2015.
Two 300~s images were obtained and then coadded in each of the $V$ and $R$ filters.
Unfortunately, the optical counterpart was not detected in either band
(see Figure \ref{Fig_keck_VI_image}). An upper limit was derived for each
coadded image. All of our optical photometry is reported in Table 1.

Spectroscopic observations of the optical afterglow of GRB~150910A were performed
with the Kast double spectrograph (Miller \& Stone 1993) on the Shane 3-m telescope
at Lick Observatory, starting $\sim 1.1$~hr after the burst (Zheng et al. 2015).
Exposures of 1200~s and 2400~s were obtained covering the 3500--10,000~\AA\
wavelength range, with the long slit at or near the parallactic angle (Filippenko 1982)
to minimize differential light losses caused by atmospheric dispersion. Spectra
were reduced using standard techniques for CCD processing and spectrum extraction,
specifically the KastShiv\footnote{https://github.com/ishivvers/TheKastShiv} pipeline.
Low-order polynomial fits to calibration-lamp spectra were used to determine the
wavelength scale, and small adjustments derived from night-sky lines in the target
frames were applied. Flux calibration and telluric-band removal were done with our own
IDL routines; details are described by Silverman et al. (2012) and Shivvers et al. (2019).

\begin{figure}
 \centering
\includegraphics[angle=0,width=0.6\textwidth]{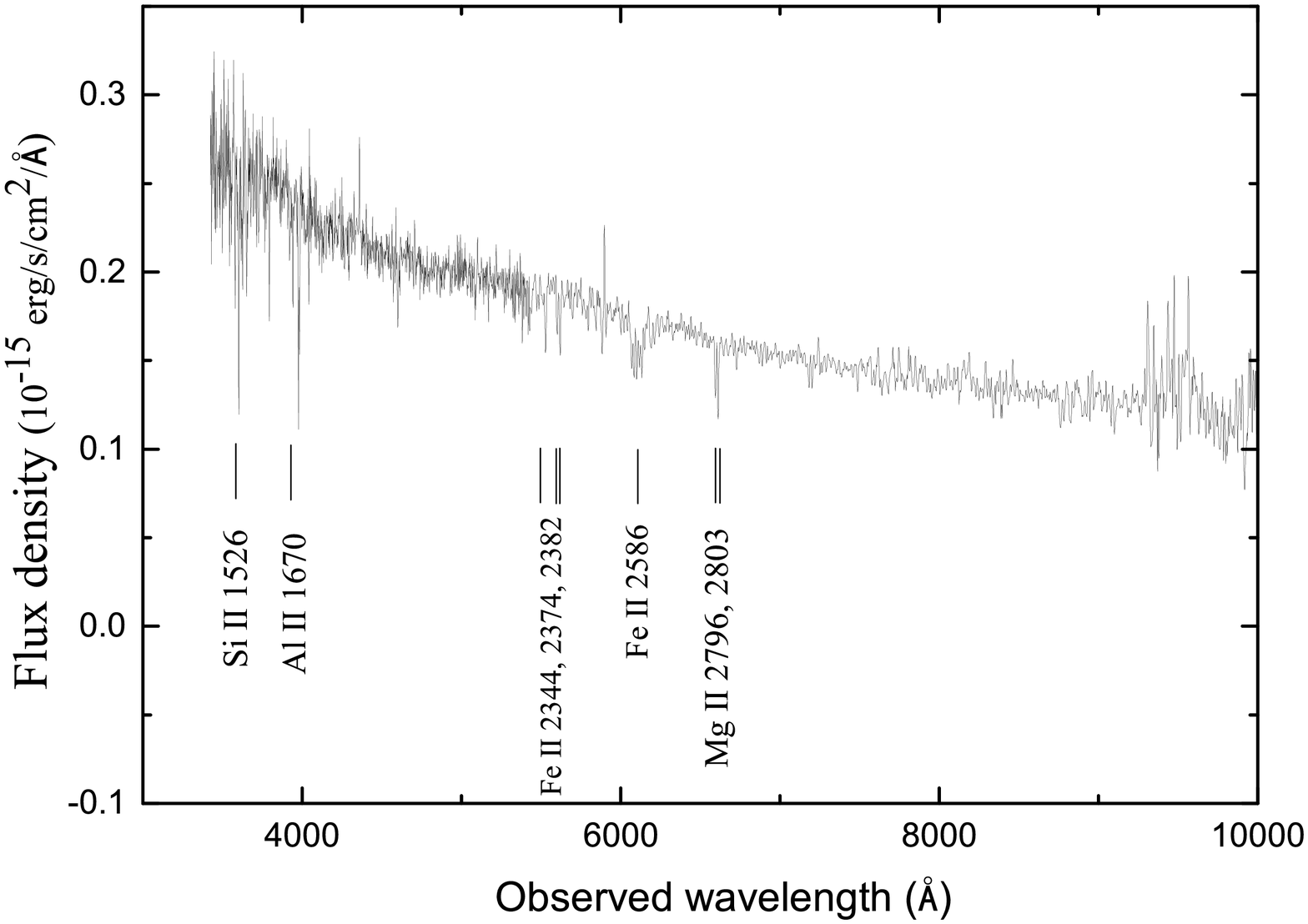}
\caption{The optical spectrum of GRB~150910A obtained with the 3-m Shane telescope at Lick Observatory.}
\label{fig-LS}
\end{figure}

The spectrum (Figure \ref{fig-LS}) exhibits a blue continuum.
We detect absorption lines from \ion{Mg}{2} $\lambda\lambda$2796, 2803 and
\ion{Fe}{2} $\lambda\lambda$2344, 2374, 2383 at a common redshift of $z=1.3585$,
as well as additional lines further to the blue (as marked in Figure \ref{fig-LS}).
We suggest this to be the redshift of the GRB.

We derived the X-ray light curve and spectrum of GRB~150910A observed with BAT. To present the optical light curve with broad temporal coverage we also include photometric data reported in the GCN Circulars (as listed in Table 1). Its XRT light curve is taken from the website of the {\em Swift} burst analyser (Evans et al. 2010)\footnote{http://www.swift.ac.uk/burst\_analyser/}. In order to make a joint X-ray light curve in the XRT band (0.3--10 keV) from the BAT trigger time to late epochs, the light curve of the prompt X-ray emission of a GRB is derived by extrapolating the BAT spectrum to the XRT band (O'Brien et al. 2006; Evans et al. 2007; Evans et al. 2009).

\section{Data Analysis}
\begin{figure}
 \centering
\includegraphics[angle=0,width=0.8\textwidth]{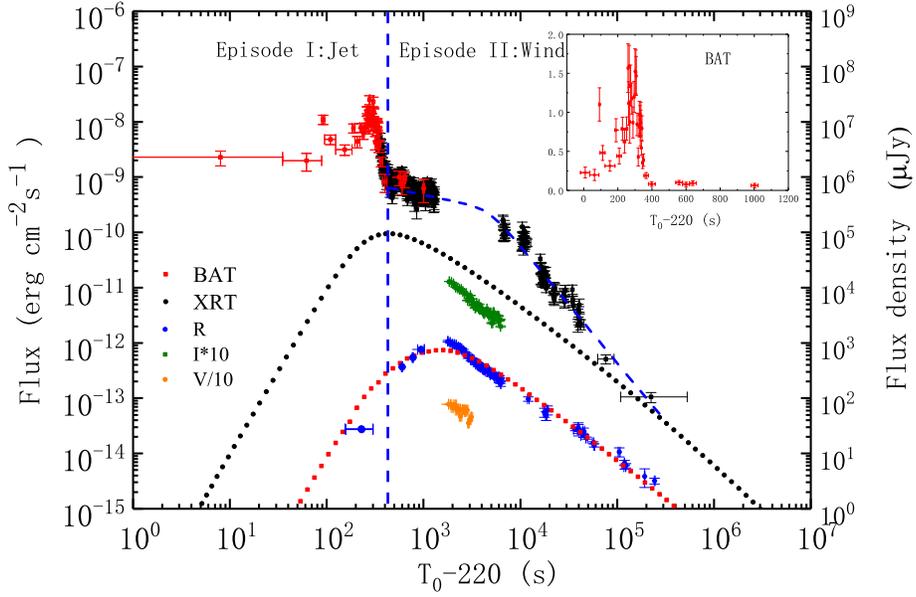}
\caption{Prompt and afterglow light curves of GRB 150910A together with our fits. Red dots are prompt X-ray data in the XRT band (0.3--10 keV) extrapolated from the gamma-ray data observed with {\em Swift}/BAT, and black dots are the XRT data. The dashed blue curve is our empirical fit to the XRT light curve with a smooth broken power-law function. The red dotted line denotes our fit to the optical data with the standard external shock model, and the black dotted line is the corresponding X-ray afterglow predicted by the external shock model. The vertical dashed line separates the jet emission (Episode I) and wind emission (Episode II) epochs based on our analysis. The inset shows the BAT light curve  on a linear scale. }
\label{fig-LCs}
\end{figure}

Figure \ref{fig-LCs} shows our optical afterglow light curve together with the X-ray light curve in the 0.3--10 keV band derived from the XRT and BAT data. Note that GRB~150910A was triggered in image-trigger mode. A weak gamma-ray signal was detected much prior to the BAT trigger time and lasted up to $\sim T_0 + 800$~s, as shown in the inset of Figure \ref{fig-LCs}. Therefore, we illustrate the joint light curves by setting a zero time of the burst at $T_0-220$~s.

The time-integrated spectrum collected with BAT from $T_0-220$ to $T_0+800$~s can be adequately fit with a single power-law function. The derived photon index is $\Gamma_{\rm I}=1.42\pm 0.12$. The fluence in the 15--150 keV band is $S_{\gamma}=(4.8\pm 0.4)\times 10^{-6}$ erg cm$^{-2}$. The peak photon energy ($E_{p}$) of the $\nu f_\nu$ spectrum should be above the BAT energy band.  We determine the $E_p$ value by using an empirical estimate as $\log E_p=(2.76\pm 0.07)-(3.61\pm 0.26)\log \Gamma_{\rm I}$ and obtain $E_p\approx 162$ keV. We take the spectral indices of the Band function $\alpha_1=-1$ and $\alpha_2=-2.3$, and make the $K$-correction for the fluence in the 1--$10^4$ keV band (e.g., Bloom et al. 2001). We find $K=1.29$. The BAT light curve peaks at $\sim T_{0}+83$~s, and the 1~s peak photon flux measured from $T+82.89$~s in the 15--150 keV band is $P=1.1\pm 0.4$ ph cm $^{-2}$ s$^{-1}$ (Pagani et al. 2015). All of the quoted uncertainties are at the 90\% confidence level. With a redshift of $z=1.36$, we obtain the burst isotropic gamma-ray energy as $E_{\gamma, iso}=(2.39\pm0.19)\times 10^{52}$ erg and a peak luminosity of $L_{\rm p, iso}=1.03\times 10^{51}$ erg s$^{-1}$.

The X-ray light curve with temporal coverage from $T_0-220$ to $T_0+800$~s derived from the data observed with BAT and XRT shows two distinct episodes. The first episode (Episode I) lasts from the beginning of the BAT observation ($T_0-220$) to $\sim T_0+140$~s, ending with a steep decay segment. The second episode (Episode II) is dominated by a long-lasting, steady emission component, which was simultaneously detected with both the BAT and XRT and rapidly decayed at around 10~ks after the BAT trigger. An empirical fit with a smooth broken power-law model of $F=F_0\left [\left ( {t}/{t_b}\right)^{\omega\alpha_1}+\left ({t}/{t_b}\right)^{\omega\alpha_2}\right]^{1/\omega}$ to the light curve of the Episode II yields $\alpha_{\rm X,1}=-0.36\pm 0.03$ and $\alpha_{\rm X,2}= -2.12\pm 0.02$, and $t_b=4.73$~ks by fixing the sharpness parameter $\omega$ at $3$ (e.g., Liang et al. 2007).

The $R$-band afterglow light curve exhibits an onset feature as predicted by the standard fireball model in the thin-shell case (Sari \& Piran 1999; Zhang et al. 2003). Such a feature was observed in about 1/3 of well-sampled optical afterglow light curves (Li et al. 2012). Our empirical fit with the smooth broken power-law function gives $\alpha_{\rm R,1}=2.24\pm0.16$ and $\alpha_{\rm R,2}=-1.36\pm0.03$, and the peak time as $t_p=1451\pm 51$~s.  One can observe that the optical afterglow light curve is completely different from the X-ray/gamma-ray light curve of the Episode II.

Comparison of the temporal slopes between the optical and X-ray light curves of GRB 150910A throughout Episode II reveals an apparent mismatch. The optical emission has a smooth onset bump followed by a normal decay, as predicted in the standard external shock model. The X-ray emission exhibits a plateau followed by a steep decay. The light curves exhibit obvious chromatic behavior with an X-ray excess (as shown in Figure \ref{fig-LCs}), which may indicate that they have different physical origins (Wang et al. 2015).

\begin{figure}
 \centering
\includegraphics[angle=0,width=0.8\textwidth]{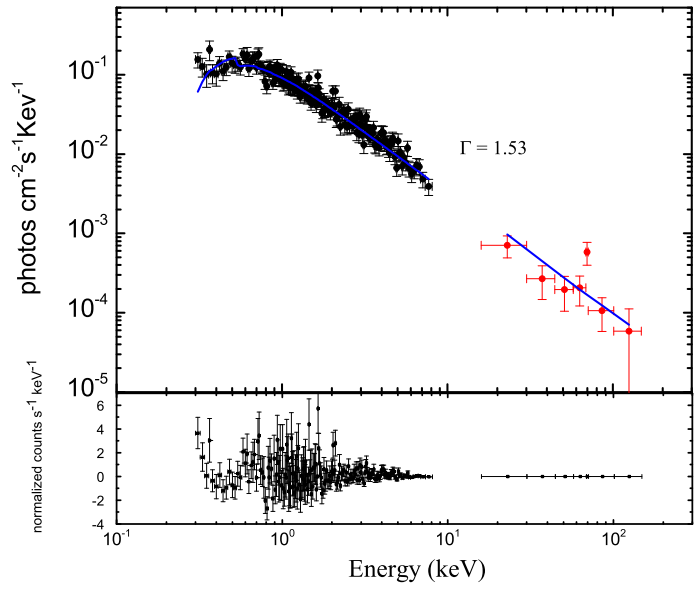}
\caption{Joint spectrum observed with BAT and XRT during the X-ray plateau in the time interval of $\{T_0+140, T_0+t_b\}$~s. The blue solid line is our best fit by a single power-law model with absorbtion.}
\label{fig-Spec}
\end{figure}

\begin{figure}
 \centering
\includegraphics[angle=0,width=0.6\textwidth]{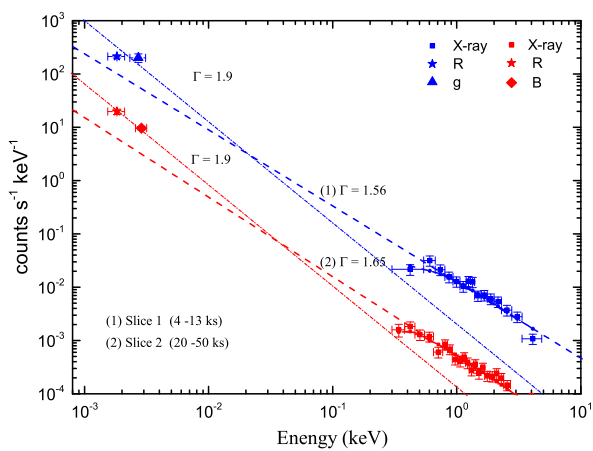}
\caption{X-ray through optical spectral energy distributions of two selected time
intervals (Slices 1 and 2). Data in the XRT (circle), $R$ (star), $B$ (diamond), and $g$ (triangle) bands are represented with different symbols. The solid lines are the best fits by an absorbed single power-law model for the XRT data only, and the dashed lines are the extrapolations of the power-law model to the optical bands. The observed optical fluxes are higher than the extrapolated values. The fitting results of photon indices from the external shock afterglow model are also plotted, along with extrapolations of the power-law model (dash-dotted lines). The data and results of Slices 1 and 2 are marked with blue and red colors, respectively.}
\label{X-ray-opt}
\end{figure}

To investigate this chromatic behavior, we perform a spectral analysis of Episode II. The joint spectrum observed with BAT and XRT from $T_0+140$ to $T_0+t_b$ is extracted. We use the Xspec package to fit the spectrum with an absorbed single power law by fixing the equivalent hydrogen column density of our Milky Way Galaxy in the burst direction as $N^{\rm MW}_{\rm H} = 5.43\times10^{20} ~\rm cm^{-2}$. We obtain an $N_{\rm H}$ value in the host galaxy as $N^{\rm host}_{\rm H} \approx (1.3\pm1.2)\times10^{21} ~\rm cm^{-2}$ and a photon index of $\Gamma=-1.53\pm 0.03$ (as shown in Figure \ref{fig-Spec}). Therefore, the X-ray and gamma-ray radiation of Episode II should be from the same component. We extract the spectra of the X-ray data in time intervals of 4--13~ks (Slice 1) and 20--50~ks (Slice 2). The X-ray spectra of the two slices can be fitted well with a power-law function with a single power law having photon indices $\Gamma$ of 1.56 and 1.65, respectively (as shown in Figure \ref{X-ray-opt} and Table 3). Extrapolating the unabsorbed power-law spectrum to the optical $B$ and $g$ bands (Rumyantsev et al. 2015; Kuroda et al. 2015a; Kuroda et al. 2015b), we find that the actual observed optical afterglows are much brighter than the extrapolation. Therefore, one cannot explain the shallow-decay X-ray flux in Episode II with energy injection to the external shocks. The physical origin of the optical and X-rays should be different.

Based on our analysis above, the X-ray and optical light curves show chromatic behavior with an excess in the X-ray flux. After the plateau, the X-ray and optical light curves did not fall at the same rate. The temporal indices of X-rays in Episode II are consistent with the prediction of a newly-born magnetar (e.g., Zhang \& M\'{e}sz\'{a}ros 2001). Therefore, we propose that GRB 150910A is typical of GRBs driven by a newly-born magnetar. The prompt gamma-rays and X-rays observed in Episode I would be from the jet radiation, and the late-time gamma-rays and X-rays observed in Episode II would be dominated by the MD wind of a newly-born magnetar via an internal energy dissipation process. The early-time optical bump might be attributed to the afterglow of the jet when it propagates into the circumburst medium.

\section{Properties of the Jet and Central Engine}
If the prompt gamma-ray emission and optical afterglow are produced from the jet, as suggested above, we can fit the optical afterglow data with the standard external shock model (Sari, Piran \& Narayan 1998; Fan \& Piran 2006). We adopt the Markov Chain Monte Carlo (MCMC) technique to evaluate the likelihoods of the model parameters. For details of model and fitting strategy, please refer to Fan et al. (2006) and Zhong et al. (2016). We fit only the optical data. The observed X-ray data place an upper limit to the X-ray afterglow.

The fitting results for GRB~150910A are illustrated in Figure \ref{fig-LCs}. Our best fit yields the following model parameters: the initial Lorentz factor of the jet $\Gamma_0=200^{+44}_{-34}$, the internal energy partitions of the electrons $\epsilon_e=(6.0\pm0.2)\times10^{-2}$ and of the magnetic field $\epsilon_B=(1.8\pm0.1)\times10^{-4}$, the circumburst medium density $n=2.57\pm0.5$ cm$^{-3}$, the isotropic kinetic jet energy $E_{\rm K,iso}=(2.0\pm0.1)\times10^{53}$ erg, and the power-law index of emitting electrons $p=2.79\pm0.07$. The efficiency of the GRB jet is $\eta_{\rm \gamma} ={E_{\rm \gamma, iso}}/({E_{\rm \gamma, iso}+E_{\rm K, iso}})=11 \%$, being similar to typical GRBs (e.g., Zhang et al. 2006). In Episode II, the photon index of the optical emission ($\Gamma = (p+1)/2 = 1.9$) is quite different from that of the X-ray emission ($\Gamma \approx  1.56$ and 1.65). Figure \ref{X-ray-opt} shows that extrapolation of the afterglow model from optical to X-rays undershoots the observed X-ray flux, consistent with the X-ray flux excess in Figure \ref{fig-LCs}).

\begin{figure}
 \centering
\includegraphics[angle=0,width=0.6\textwidth]{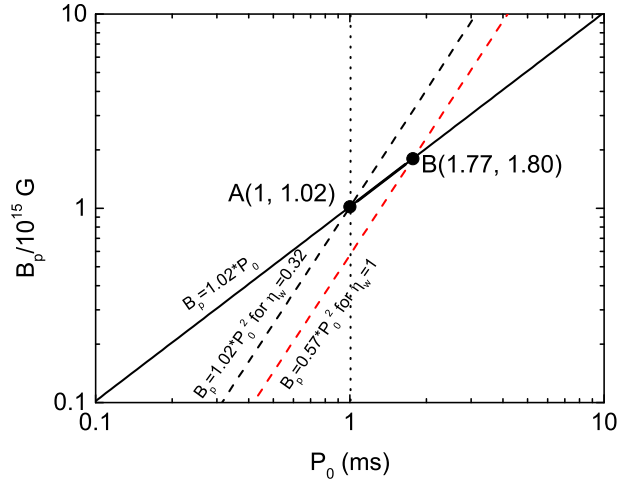}
\caption{Initial spin period $P_0$ vs. surface polar cap magnetic field strength $B_p$ distributions, which are constrained by the radiation efficiency of the magnetic dipole wind. The black and red dashed lines correspond to efficiencies of 32\% and 100\%, respectively. The vertical black dotted line is the lower limit of the spin period of a neutron star (Lattimer \& Prakash 2004). The labels ``A'' and ``B'' indicate the lower and upper limits of ($P_0$, $B_p$) with (1~ms, $1.02 \times 10^{15}$~G) and (1.77~ms, $1.80 \times 10^{15}$~G), respectively. The range between A to B is the available parameter space for GRB 150910A.}
\label{Bp-P0}
\end{figure}

The injected kinetic luminosity to the MD wind from the spin-down of a magnetar evolves as $L_{\rm k}\propto (1+t/\tau)^{-\alpha}$, where $\tau$ is the characteristic spin-down timescale of the magnetar (e.g., Zhang \& M\'{e}sz\'{a}ros 2001). The $\alpha$ value depends on the spin-down energy lost via the MD wind or the GW radiation: $\alpha=1$ if the rotation energy loss is dominated by GW radiation, $\alpha=2$ if MD radiation dominates. The X-ray emission of Episode II may be attributed to radiation from the MD wind of a newly-born magnetar. It is consistent with the evolution of the injected kinetic luminosity from the spin-down of a newly born magnetar in the case that the spin-down energy lost is dominated by electromagnetic emission --- that is, $L_k\propto (1+t/\tau)^{-2}$. We estimate the initial spin period ($P_0$) and the surface polar cap magnetic field strength ($B_p$) of the magnetar in GRB 150910A (e.g., Zhang \& M\'{e}sz\'{a}ros 2001),
\begin{eqnarray}
B_{p,15} = 2.05(I_{45} R_6^{-3} (L_{b,49}/\eta_w)^{-1/2} \tau_{3}^{-1})~\rm G,\\
P_{0,-3} = 1.42(I_{45}^{1/2} (L_{b,49}/\eta_w)^{-1/2} \tau_{3}^{-1/2})~\rm s,
\label{Bp-tau}
\end{eqnarray}
where $R$ and $I$ are respectively the neutron star radius and moment of inertia, $\eta_w$ is the radiative efficiency of the MD wind, and the convention $Q=10^x\,Q_x$ is adopted in cgs units. One can infer the relations $B_{p}-P_0$ and $B_{p}-P_{0}^{2}$,
\begin{eqnarray}
B_{p,15}=1.44 \, I^{1/2}_{45}R^{-3}_{6} \tau^{-1/2}_{3} P_{0,-3}~{\rm G}, \\
B_{p,15}=1.02 \, R^{-3}_{6} (L_{b,49}/\eta_w)^{1/2} P^{2}_{0,-3}~{\rm G}.
\end{eqnarray}
Our above analysis yields $\tau=t_b/(1+z)=2007$~s and $L_b=3.19\times 10^{48}$ erg s$^{-1}$. By taking $I_{45}=1$, $R_6=1$, and a lower limit of $P_0$ for a neutron star as $P_{0,-3}\gtrsim 1$ (e.g., Lattimer \& Prakash 2004), we have $B_{p, 15}\geq 1.02$ and $\eta_w\geq 32\%$ (point A in Figure \ref{Bp-P0}). Since $\eta_w\leq 1$, we also have $B_{p, 15}\leq 1.80$ and $P_{0,-3} \leq 1.77$ (point B in Figure \ref{Bp-P0}). Thus, we obtain tight constraints on $B_p$ and $P_0$ as $1.02 \leq B_{p, 15} \leq 1.80$ and $1\leq P_{0, -3}\leq 1.77$ (the range between points A and B in Figure \ref{Bp-P0}).

\section{Discussion}
\begin{figure}
 \centering
\includegraphics[angle=0,width=0.6\textwidth]{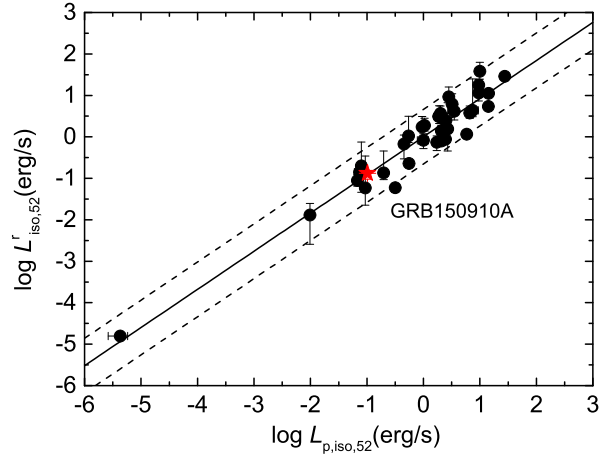}
\caption{GRB~150910A (marked with a red star) shares the some empirical relation $L_{\rm p,iso}$ -- $E_{\rm p,z}$ -- $\Gamma_0$ with typical GRBs (black circles; data from Liang et al. 2015). The solid and dashed lines are the least-squares fit and its 95\% confidence levels, respectively. $L_{\rm p,iso}$ is the peak of the isotropic luminosity, and $L^r_{\rm iso}$ is derived from three-parameter correlations $L_{\rm p,iso}$ -- $E_{\rm p,z}$ -- $\Gamma_0$ by Liang et al. (2015).}
\label{jet-}
\end{figure}

Our analysis shows that the optical observations of GRB 150910A are well explained with the external shock model. The early-time optical bump is then attributed to the deceleration of the jet by the ambient medium. Such a feature may also be interpreted with the line-of-sight effect for a uniform jet with a sharp edge (Panaitescu \& Vestrand 2008; Guidorzi et al. 2009; Margutti et al. 2010). In this scenario, the optical light curve may peak at a time when the jet Lorentz factor satisfies $\Gamma=1/(\theta_v-\theta_j)$, where $\theta_v$ and $\theta_j$ are the viewing angle and the jet opening angle, respectively. By analysing a sample of optical light curves with an onset bump feature, Liang et al. (2010) argued that such a feature would result from the jet deceleration and that $\Gamma_0$ of the jet should be robustly estimated with the peak time of the optical bump. We examine whether GRB 150910A follows the same empirical $L_{\rm p,iso}-E_{\rm p,z}-\Gamma_0$ relation determined for typical GRBs (Liang et al. 2015), where $E_{\rm p,z}$ is the peak energy in the cosmological rest frame. Figure \ref{jet-} illustrates this consistency, likely suggesting that the derived $\Gamma_0$ is the initial Lorentz factor of the fireball and the onset bump may be due to the deceleration of the fireball, as in typical long GRBs.

\begin{figure}
 \centering
\includegraphics[angle=0,width=0.43\textwidth]{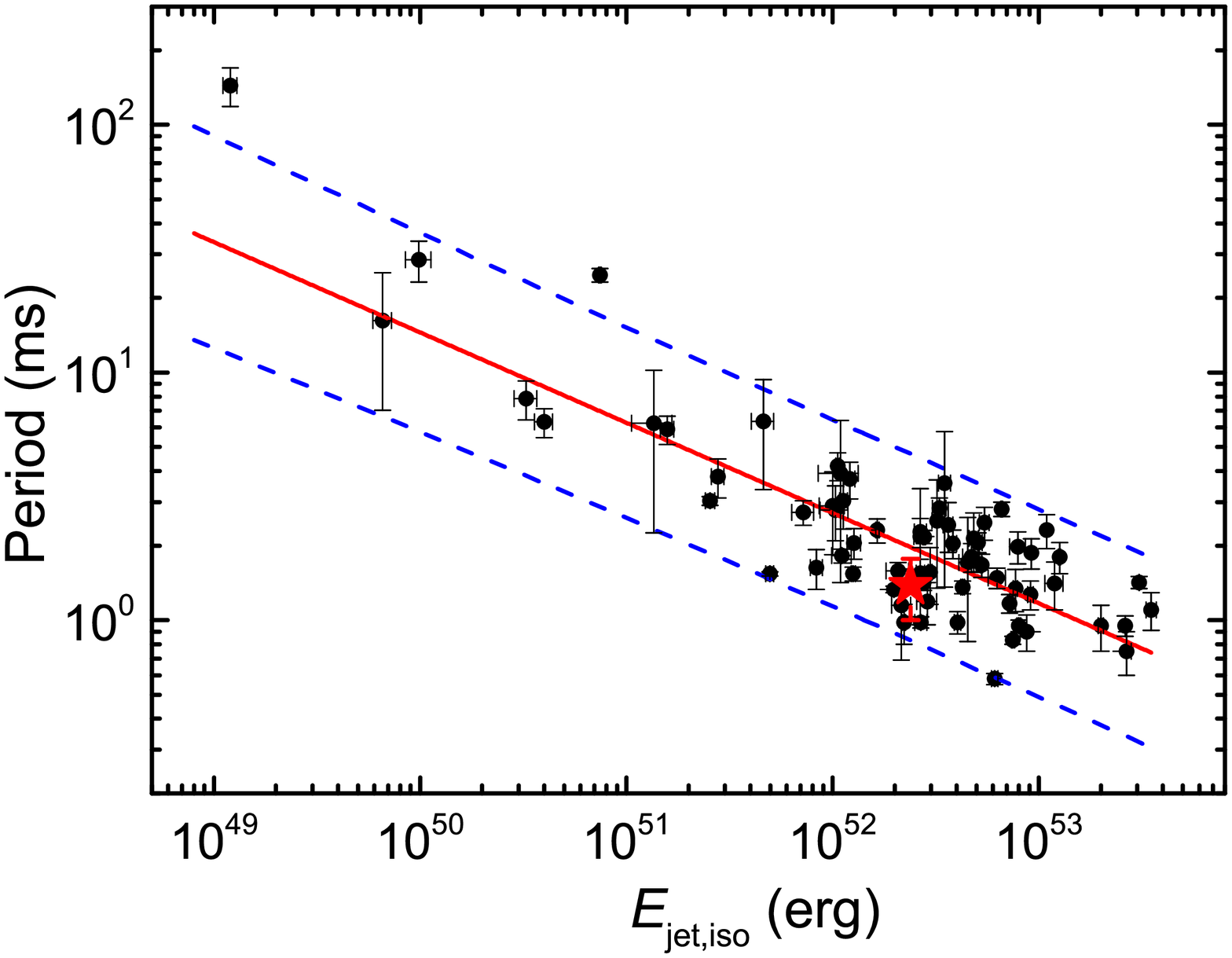}
\includegraphics[angle=0,width=0.42\textwidth]{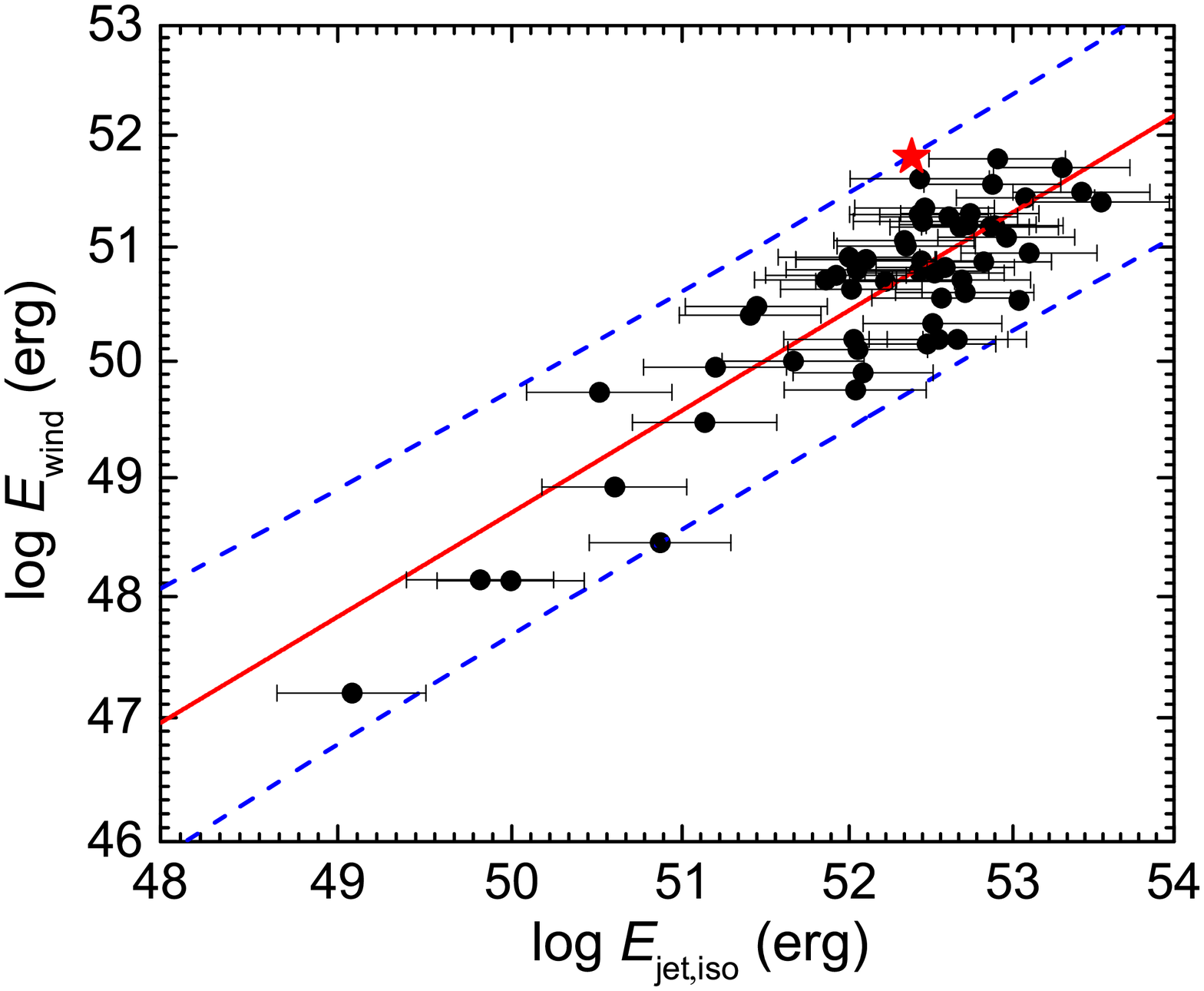}

\caption{The correlations of $P_0$ -- $E_{\rm jet, iso}$ (left panel) and $E_{\rm wind}$ -- $E_{\rm jet, iso}$ (right panel); GRB~150910A is marked with a red star. The solid and dashed lines are the least-squares fit and its 95\% confidence levels, respectively. The sample of Zou et al. (2019) (black point) also presented here.}
\label{jet-wind}
\end{figure}

The X-ray plateau of GRB 150910A was simultaneously observed in the BAT and XRT bands. We calculate the energy of the MD wind as $E_{\rm wind}=L_{\rm wind}\times \tau\approx 6.40\times 10^{51}$ erg, where $L_{\rm wind}$ is the observed wind luminosity in the BAT+XRT band. Figure \ref{jet-wind} shows GRB 150910A in the $E_{\rm wind}-E_{\rm jet}$ plane in comparison with a sample of GRBs whose early XRT light curves are dominated by MD radiation (Zou et al. 2019). One can observe that the MD wind of GRB 150910A is the most energetic one among these GRBs. However, it still follows the $P_0-E_{\gamma, jet}$ relation reported by Zou et al. (2019).

Such a jet-wind coexisting system may explain the observed diverse temporal features in the optical and X-ray afterglow light curves. Comprehensive analysis of both the optical and X-ray afterglow light curves reveals that the light-curve diversity may be due to the competition among radiation components (e.g., Li et al. 2012; Liang et al. 2013). The optical afterglows and the single power-law decaying X-ray afterglows may be dominated by the jet afterglows, and the X-ray emission in the shallow-decaying segment of the canonical XRT light curves may be dominated by MD radiation (e.g., Zou et al. 2019). GRB 150910A is unusual with its extremely energetic wind, which overwhelmingly dominates the observed early X-ray plateau. The MD radiation decays as roughly $L_k\propto t^{-2}$ after the characteristic spin-down timescale $\tau$, and $\tau$ is typically thousands of seconds. In addition, the X-ray afterglow usually decays as roughly $L_{a}\propto t^{-1.2}$ prior to the jet break. The observed jet-break time is usually at several days (e.g., Liang et al. 2008). Therefore, the observed X-ray emission at late epoch (several hours after the GRB trigger) may be dominated by the jet afterglow, where they will have the same decay slopes in both X-ray and optical light curves. Wang et al. (2015) found that a large fraction of optical and X-ray afterglows can still be explained with the external shock model. For GRB 150910A, the temporal slopes of the X-ray and optical light curves fall at the same rate until after $10^5$~s (as shown in Figure \ref{fig-LCs}).
\section{Conclusions}
\label{sec:Conclusion}
We report our optical spectroscopic and photometric observations of the optical afterglow of GRB~150910A, and we investigate the physical origins of both the optical and X-ray afterglows, incorporating data obtained with the {\it Swift} BAT and XRT. We show that the gamma-ray and X-ray emission of this GRB can be separated into the jet-emission episode (Episode I) and the magnetar MD wind radiation episode (Episode II). The jet-emission episode is observed with BAT more than 200~s prior to its trigger.

Modeling the $R$-band optical light curve with the standard external shock model, we obtain jet parameters of $\Gamma_0=200^{+44}_{-34}$, $\epsilon_e=0.06\pm 0.002$, $\epsilon_B=(1.8\pm0.1)\times10^{-4}$, $n=2.57\pm 0.5$ cm$^{-3}$, and $E_{\rm K, iso}=(2.0\pm 0.1)\times 10^{53}$ erg. The radiative efficiency of the jet prompt emission is $\eta_{\gamma}\approx 11\%$. This GRB follows the $L_{\rm p,iso}-E_{\rm p,z}-\Gamma_0$ relation derived for typical GRBs that have a clear detection of an onset bump in their early-time optical afterglow light curves.

The MD wind emission was detected in both the BAT and XRT bands, making GRB~150910A the brightest among the current sample of MD winds detected by the XRT. We infer the parameters of the magnetar as 1.02$\times 10^{15}$ G $\leq B_{p} \leq 1.80\times 10^{15}$ G and 1 ms $\leq P_{0}\leq 1.77$ ms, and the lower limit of the radiation efficiency of the wind as $\eta_w\geq32\%$. It also satisfies the $P_0-E_{\rm jet}$ relation of GRBs in which a shallow decay segment was detected in their early-time XRT light curves.

\section*{ Acknowledgements }

We are grateful to Daniel Perley for the Lick/Kast observation and data reduction, Minkyu Kim and Timothy William Ross for help obtaining the Lick/Nickel images, and Melissa L. Graham for her assistance at Keck. Wei-Hua Lei, Zi-Gao Dai, and Bing Zhang are acknowledged for helpful discussions. This research used public data from the {\it Swift} data archive and the UK {\it Swift} Science Data Center. This work is supported by the National Natural Science Foundation of China (grants 11673006, 11773007, and 11533003), the Guangxi Science Foundation (grant 2016GXNSFFA380006, 2017AD22006, 2018GXNSFFA281010, 2016GXNSFDA380027, and 2018GXNSFDA281033), the One-Hundred-Talents Program of Guangxi Colleges, and the High Level Innovation Team and Outstanding Scholar Program in Guangxi Colleges. A.V.F.'s supernova/GRB group is grateful for financial assistance from the TABASGO Foundation, the Christopher R. Redlich Fund, NASA/{\it Swift} grants NNX12AD73G and 80NSSC19K0156, and the Miller Institute for Basic Research in Science (U.C. Berkeley). KAIT and its ongoing operation were made possible by donations from Sun Microsystems, Inc., the Hewlett-Packard Company, AutoScope Corporation, Lick Observatory, the U.S. National Science Foundation, the University of California, the Sylvia \& Jim Katzman Foundation, and the TABASGO Foundation. Research at Lick Observatory is partially supported by a generous gift from Google. Some of the data presented herein were obtained at the W. M. Keck Observatory, which is operated as a scientific partnership among the California Institute of Technology, the University of California, and NASA; the observatory was made possible by the generous financial support of the W. M. Keck Foundation.

\clearpage

\begin{deluxetable}{cccccc}
\tabletypesize{\small}
\tablewidth{0pt}
\label{Tab:data}
\tablecaption{Optical Afterglow Photometry Log of GRB 150910A.}
\tablehead{
\colhead{$T-T_0$(mid, s)} &
\colhead{Exp (s)} &
\colhead{Mag$^{a}$} &
\colhead{$\sigma^{a}$} &
\colhead{Filter} &
\colhead{Telescope, GCN Circ., Ref.}
}
\startdata
\object{	228  	}&	 73    &	19.90	&	0.01	&	$W$	&	UVOT,18270,(1)\\
\object{	229     }&	 72    &	19.93	&	0.01	&	$W$	&	UVOT,18270,(1)\\
\object{	602  	}&	 9.5   &	17.09	&	0.01	&	$W$	&	UVOT,18270,(1)\\
\object{	775  	}&	 9.5   &	16.67	&	0.01	&	$W$	&	UVOT,18270,(1)\\
\object{	941  	}&	 73    &	16.30	&	0.01	&	$W$	&	UVOT,18270,(1)\\
\object{	578  	}&	 9.5   &	17.56	&	0.01	&	$b$	&	UVOT,18270,(1)\\
\object{	751  	}&	 9.5   &	16.87	&	0.01	&	$b$	&	UVOT,18270,(1)\\
\object{	1815    }&	 10	   &	16.81	&	0.12	&	$V$	&	KAIT\\
\object{	1915	}&	 10	   &	16.82	&	0.13	&	$V$	&	KAIT\\
\object{	2014	}&	 10	   &	16.96	&	0.14	&	$V$	&	KAIT\\
\object{	2112	}&	 10	   &	16.83	&	0.12	&	$V$	&	KAIT\\
\object{	2212	}&	 10	   &	17.08	&	0.16	&	$V$	&	KAIT\\
\object{	2312	}&	 10    &	17.01	&	0.15	&	$V$	&	KAIT\\
\object{	2412    }&	 10	   &	17.36	&	0.17	&	$V$	&	KAIT\\
\object{	2510	}&	 10	   &	17.05	&	0.17	&	$V$	&	KAIT\\
\object{	2614	}&	 10	   &	17.06	&	0.18	&	$V$	&	KAIT\\
\object{	2714	}&	 10	   &	17.15	&	0.17	&	$V$	&	KAIT\\
\object{	2814	}&	 10	   &	17.07	&	0.12	&	$V$	&	KAIT\\
\object{	2914	}&	 10    &	17.66	&	0.14	&	$V$	&	KAIT\\
\object{	3013    }&	 10	   &	17.52	&	0.20	&	$V$	&	KAIT\\
\object{	3113	}&	 10	   &	17.38	&	0.22	&	$V$	&	KAIT\\
\object{	2,583,936	}&	 $2 \times 300$	 &	$>$24.30	&	---	&	$V$	&	Keck\\
\object{	1849    }&	 10	   &	15.74	&	0.08	&	$I$	&	KAIT\\
\object{	1947	}&	 10	   &	15.78	&	0.08	&	$I$	&	KAIT\\
\object{	2045	}&	 10	   &	15.85	&	0.08	&	$I$	&	KAIT\\
\object{	2145	}&	 10	   &	15.99	&	0.09	&	$I$	&	KAIT\\
\object{	2245	}&	 10	   &	16.04	&	0.09	&	$I$	&	KAIT\\
\object{	2345	}&	 10    &	16.10	&	0.12	&	$I$	&	KAIT\\
\object{	2445	}&	 10	   &	16.10	&	0.09	&	$I$	&	KAIT\\
\object{	2549	}&	 10	   &	16.19	&	0.12	&	$I$	&	KAIT\\
\object{	2648	}&	 10	   &	16.27	&	0.08	&	$I$	&	KAIT\\
\object{	2748	}&	 10	   &	16.28	&	0.10	&	$I$	&	KAIT\\
\object{	2848	}&	 10	   &	16.60	&	0.13	&	$I$	&	KAIT\\
\object{	2946	}&	 10	   &	16.51	&	0.11	&	$I$	&	KAIT\\
\object{	3046	}&	 10	   &	16.38	&	0.09	&	$I$	&	KAIT\\
\object{	3144	}&	 10	   &	16.54	&	0.11	&	$I$	&	KAIT\\
\object{	3209	}&	 10	   &	16.67	&	0.16	&	$I$	&	KAIT\\
\object{	3276	}&	 10	   &	16.82	&	0.13	&	$I$	&	KAIT\\
\object{	3342	}&	 10	   &	16.76	&	0.11	&	$I$	&	KAIT\\
\object{	3409	}&	 10	   &	16.95	&	0.14	&	$I$	&	KAIT\\
\object{	3476	}&	 10	   &	16.89	&	0.12	&	$I$	&	KAIT\\
\object{	3541	}&	 10	   &	16.74	&	0.13	&	$I$	&	KAIT\\
\object{	3608	}&	 10	   &	16.86	&	0.14	&	$I$	&	KAIT\\
\object{	3674	}&	 10	   &	16.93	&	0.13	&	$I$	&	KAIT\\
\object{	3739	}&	 10	   &	16.78	&	0.13	&	$I$	&	KAIT\\
\object{	3806	}&	 10	   &	17.01	&	0.14	&	$I$	&	KAIT\\
\object{	3873	}&	 10	   &	16.97	&	0.13	&	$I$	&	KAIT\\
\object{	3940	}&	 10	   &	16.97	&	0.19	&	$I$	&	KAIT\\
\object{	4007	}&	 10	   &	16.99	&	0.17	&	$I$	&	KAIT\\
\object{	4080	}&	 10	   &	17.09	&	0.11	&	$I$	&	KAIT\\
\object{	4146	}&	 10	   &	17.17	&	0.15	&	$I$	&	KAIT\\
\object{	4213	}&	 10	   &	17.15	&	0.11	&	$I$	&	KAIT\\
\object{	4280	}&	 10	   &	17.20	&	0.14	&	$I$	&	KAIT\\
\object{	4516	}&	 10	   &	16.99	&	0.16	&	$I$	&	KAIT\\
\object{	4583	}&	 10	   &	17.16	&	0.19	&	$I$	&	KAIT\\
\object{	4650	}&	 10	   &	17.11	&	0.15	&	$I$	&	KAIT\\
\object{	4717	}&	 10	   &	17.29	&	0.17	&	$I$	&	KAIT\\
\object{	4783	}&	 10	   &	17.21	&	0.16	&	$I$	&	KAIT\\
\object{	4850	}&	 10	   &	17.25	&	0.15	&	$I$	&	KAIT\\
\object{	4915	}&	 10	   &	17.16	&	0.11	&	$I$	&	KAIT\\
\object{	4982	}&	 10	   &	17.31	&	0.16	&	$I$	&	KAIT\\
\object{	5048	}&	 10	   &	17.62	&	0.21	&	$I$	&	KAIT\\
\object{	5115	}&	 10	   &	17.35	&	0.24	&	$I$	&	KAIT\\
\object{	5182	}&	 10	   &	17.62	&	0.18	&	$I$	&	KAIT\\
\object{	5249	}&	 10	   &	17.14	&	0.26	&	$I$	&	KAIT\\
\object{	5316	}&	 10	   &	17.34	&	0.19	&	$I$	&	KAIT\\
\object{	5382	}&	 10	   &	17.16	&	0.14	&	$I$	&	KAIT\\
\object{	5845	}&	 10	   &	17.30	&	0.21	&	$I$	&	KAIT\\
\object{	5912	}&	 10	   &	17.48	&	0.16	&	$I$	&	KAIT\\
\object{	5979	}&	 10	   &	17.31	&	0.16	&	$I$	&	KAIT\\
\object{	6046	}&	 10	   &	17.52	&	0.20	&	$I$	&	KAIT\\
\object{	6112	}&	 10	   &	17.48	&	0.22	&	$I$	&	KAIT\\
\object{	6177	}&	 10	   &	17.47	&	0.24	&	$I$	&	KAIT\\
\object{	6244	}&	 10	   &	17.78	&	0.30	&	$I$	&	KAIT\\
\object{	2583936	}&	 2$\times$300	 &	$>$23.30	&	---	&	$I$	&	Keck\\
\object{	1782    }&	 10	   &	16.18	&	0.04	&	$Clear$	&	KAIT\\
\object{	1882	}&	 10	   &	16.24	&	0.04	&	$Clear$	&	KAIT\\
\object{	1980	}&	 10	   &	16.31	&	0.04	&	$Clear$	&	KAIT\\
\object{	2078	}&	 10	   &	16.34	&	0.03	&	$Clear$	&	KAIT\\
\object{	2178	}&	 10	   &	16.36	&	0.04	&	$Clear$	&	KAIT\\
\object{	2278	}&	 10    &	16.45	&	0.04	&	$Clear$	&	KAIT\\
\object{	2378	}&	 10	   &	16.47	&	0.04	&	$Clear$	&	KAIT\\
\object{	2478	}&	 10	   &	16.52	&	0.03	&	$Clear$	&	KAIT\\
\object{	2581	}&	 10	   &	16.60	&	0.04	&	$Clear$	&	KAIT\\
\object{	2681	}&	 10	   &	16.70	&	0.04	&	$Clear$	&	KAIT\\
\object{	2781	}&	 10	   &	16.77	&	0.05	&	$Clear$	&	KAIT\\
\object{	2881	}&	 10	   &	16.91	&	0.04	&	$Clear$	&	KAIT\\
\object{	2979    }&	 10	   &	16.88	&	0.06	&	$Clear$	&	KAIT\\
\object{	3079	}&   10	   &	17.00	&	0.07	&	$Clear$	&	KAIT\\
\object{	3177	}&	 10	   &	16.99	&	0.05	&	$Clear$	&	KAIT\\
\object{	3242	}&	 10	   &	17.05	&	0.06	&	$Clear$	&	KAIT\\
\object{	3309	}&	 10	   &  	17.07	&	0.04	&	$Clear$	&	KAIT\\
\object{	3376	}&	 10	   &	17.12	&	0.05	&	$Clear$	&	KAIT\\
\object{	3442	}&	 10	   &	17.19	&	0.06	&	$Clear$	&	KAIT\\
\object{	3509	}&	 10	   &	17.18	&	0.06	&	$Clear$	&	KAIT\\
\object{	3574	}&	 10	   &	17.18	&	0.05	&	$Clear$	&	KAIT\\
\object{	3641	}&	 10	   &	17.27	&	0.06	&	$Clear$	&	KAIT\\
\object{	3706    }&	 10	   &	17.20	&	0.05	&	$Clear$	&	KAIT\\
\object{	3773	}&	 10	   &	17.28	&	0.05	&	$Clear$	&	KAIT\\
\object{	3840	}&	 10	   &	17.27	&	0.06	&	$Clear$	&	KAIT\\
\object{	3907	}&	 10	   &	17.37	&	0.07	&	$Clear$	&	KAIT\\
\object{	3973	}&	 10	   &	17.32	&	0.06	&	$Clear$	&	KAIT\\
\object{	4046	}&	 10	   &	17.42	&	0.05	&	$Clear$	&	KAIT\\
\object{	4113	}&	 10	   &	17.46	&	0.06	&	$Clear$	&	KAIT\\
\object{	4180	}&	 10	   &	17.45	&	0.06	&	$Clear$	&	KAIT\\
\object{	4247	}&	 10	   &	17.49	&	0.06	&	$Clear$	&	KAIT\\
\object{	4313	}&	 10	   &	17.48	&	0.06	&	$Clear$	&	KAIT\\
\object{	4340	}&	 10	   &	17.46	&	0.06	&	$Clear$	&	KAIT\\
\object{	4550	}&	 10	   &	17.55	&	0.06	&	$Clear$	&	KAIT\\
\object{	4617	}&	 10	   &	17.53	&	0.07	&	$Clear$	&	KAIT\\
\object{	4683	}&	 10	   &	17.51	&	0.06	&	$Clear$	&	KAIT\\
\object{	4750	}&	 10	   &	17.50	&	0.06	&	$Clear$	&	KAIT\\
\object{	4817	}&	 10	   &	17.60	&	0.06	&	$Clear$	&	KAIT\\
\object{	4883	}&	 10	   &	17.70	&	0.07	&	$Clear$	&	KAIT\\
\object{	4948	}&	 10	   &	17.58	&	0.06	&	$Clear$	&	KAIT\\
\object{	5015	}&	 10	   &	17.66	&	0.08	&	$Clear$	&	KAIT\\
\object{	5082	}&	 10	   &	17.65	&	0.06	&	$Clear$	&	KAIT\\
\object{	5148	}&	 10	   &	17.64	&	0.05	&	$Clear$	&	KAIT\\
\object{	5215	}&	 10	   &	17.68	&	0.06	&	$Clear$	&	KAIT\\
\object{	5282	}&	 10	   &	17.71	&	0.04	&	$Clear$	&	KAIT\\
\object{	5349	}&	 10	   &	17.76	&	0.07	&	$Clear$	&	KAIT\\
\object{	5414	}&	 10	   &	17.78	&	0.05	&	$Clear$	&	KAIT\\
\object{	5879	}&	 10	   &	17.87	&	0.07	&	$Clear$	&	KAIT\\
\object{	5946	}&	 10	   &	17.81	&	0.08	&	$Clear$	&	KAIT\\
\object{	6012	}&	 10	   &	17.99	&	0.07	&	$Clear$	&	KAIT\\
\object{	6079	}&	 10	   &	17.81	&	0.06	&	$Clear$	&	KAIT\\
\object{	6144	}&	 10	   &	17.93	&	0.09	&	$Clear$	&	KAIT\\
\object{	6211	}&	 10	   &	18.03	&	0.07	&	$Clear$	&	KAIT\\
\object{	6278	}&	 10	   &	18.11	&	0.11	&	$Clear$	&	KAIT\\
\object{	12,104	}&	270	   &	18.40	&	0.10	&	$R$	&	MITSuME,18267,(2)\\
\object{	17,676	}&	300	   &	19.40	&	0.20    &	$R$	&	Nanshan,18269,(3)\\
\object{	18,504	}&	300	   &	19.55	&	0.20	&	$R$	&	Nanshan,18269,(3)\\
\object{	19,116	}&	300	   &	19.30	&	0.20	&	$R$	&	Nanshan,18269,(3)\\
\object{	36,985	}&	1650   &	20.18	&	0.05	&	$R$	&	TSHAO,18281,(4)\\
\object{	39,492	}&	150    &	20.09	&	0.23	&	$R$	&	T100,18314,(5)\\
\object{	39,726	}&	900    &	20.20	&	0.05	&	$R$	&	SAO RAS,18275,(6)\\
\object{	39,816	}&	150	   &	20.18	&	0.25	&	$R$	&	T100,18314,(5)\\
\object{	44,905	}&	2760   &	20.46	&	0.07	&	$R$	&	Mt-Terkol,18306,(7)\\
\object{	45,131	}&	990	   &	20.25	&	0.14	&	$R$	&   Chuguev,18287,(8)\\
\object{	48,412	}&	900	   &	20.57	&	0.05	&	$R$	&	SAO RAS,18275,(6)\\
\object{	56,660	}&	900	   &	20.78	&	0.05	&	$R$	&	SAO RAS,18275,(6)\\
\object{	57,959	}&	780	   &	20.90	&	0.07    &	$R$	&	Mt-Terkol,18320,(9)\\
\object{	78,752	}&	 5$\times$600	   &	21.17	&	0.09	&	$R$	&	Nickel\\
\object{	104,998	}&	2400   &	21.80	&	0.20	&	$R$	&	MITSuME,18288,(10)\\
\object{	118,903	}&	600    &	21.69	&	0.12	&	$R$	&	CrAO,18556,(11)\\
\object{	119,882	}&	2400   &	21.87	&	0.10	&	$R$	&	TSHAO,18319,(12)\\
\object{	123,068	}&	600	   &	21.82	&	0.10	&	$R$	&	CrAO,18556,(9)\\
\object{	192,376	}&	1620   &	22.30	&	0.40	&	$R$	&	TSHAO,18319,(12)\\
\object{	121,150	}&	600    &	22.48	&	0.15	&	$B$	&	CrAO,18556,(9)\\
\object{	124,727	}&	960    &	22.09	&	0.10	&	$B$	&	CrAO,18556,(9)\\
\enddata
\tablecomments{To complete our analysis, we adopt additional photometric data published in the GCN Circulars listed below.\\
(a) Not corrected for Galactic foreground reddening.\\
The reference time $T_0$ is the {\em Swift} BAT burst trigger time.\\
``$T-T_0$" is the middle time (s) for each observation.\\
``Exposure" is the exposure time (s) for each observation. \\
``$\sigma$" means the uncertainty in the magnitude.\\
References: (1) McCauley \& Pagani (2015); (2) Kuroda et al. (2015a); (3) Xu et al. (2015); (4) Mazaeva et al. (2015); (5) Sonbas et al. (2015); (6) Moskvitin \& Goranskij (2015); (7) Andreev et al. (2015); (8) Krugly et al. (2015); (9) Volnova et al. (2015a); (10) Kuroda et al. (2015b); (11) Rumyantsev et al. (2015); (12) Volnova et al. (2015b).
}

\end{deluxetable}

\begin{table}
\caption{Results of Empirical Fits with a Smooth Broken Power-Law Function.}
\centering
\tabletypesize{\footnotesize}
\begin{tabular}{cccccccc}
\hline\hline
Band & $F_{\rm 0} (\rm erg\ cm^{-2}\ s^{-1})$& $\alpha_1$ & $\alpha_{2}$ &  $t_{b} (s)$ &  $\chi^2/{\rm dof}$\\
\hline
Optical  & $(1.02\pm0.05)\times10^{-11}$ & $2.44\pm0.27$ &  $-1.32\pm0.01$  & $1095\pm39$  & 1.50 \\
X-ray  & $(2.71\pm0.26)\times10^{-10}$  &  $-0.36\pm0.03$  &  $-2.12\pm0.02$ & $4518\pm332$  & 3.01 \\
\hline
\end{tabular}
\label{Tab2}
\end{table}

\begin{table}
\tabletypesize{\footnotesize}
\tablewidth{500pt}
\caption{Spectral Analysis of Afterglow Emission.}
\centering
\begin{tabular}{cllc}
\hline\hline
  Slice  & Interval & $\chi_r^{2}$ &  $\Gamma$   \\
\hline
   1   &       4--13 (ks)& $1.50$   &  $-1.56\pm0.01$    \\
   2   &     20--50  (ks)& $0.96$   &   $-1.65\pm0.01$    \\

\hline
\end{tabular}
\label{Tab3}
\end{table}

\clearpage

\end{document}